\documentclass[aps,showpacs,preprintnumbers,amsmath, amssymb]{revtex4}

\usepackage{float}
\usepackage{graphics,epsfig}
\usepackage{graphicx}
\usepackage{dcolumn}
\usepackage{bm}

\begin{document}
\baselineskip=0.8 cm
\title{{\bf Holographic insulator/superconductor transitions in the three dimensional AdS soliton}}
\author{Yan Peng$^{1,2}$\footnote{yanpengphy@163.com}}
\affiliation{\\$^{1}$ School of Mathematical Sciences,
 Qufu Normal University, Qufu, Shandong 273165, P. R. China
\\$^{2}$ School of Mathematics and Computer Science,
\\Shaanxi University of Technology, Hanzhong, Shaanxi 723000, P. R. China  }

\vspace*{0.2cm}
\begin{abstract}
\baselineskip=0.6 cm
\begin{center}
{\bf Abstract}
\end{center}

We investigate the holographic description of a superconductor
constructed in the (2+1)-dimensional AdS soliton background in the probe limit.
We study the holographic properties through both analytical and numerical methods.
With analytical methods, we are the first to obtain the exact formula for critical
phase transition points as $\mu_{c}=1+\sqrt{1+m^2}$.
Around the transition points, we find a correspondence between the
value of the scalar field at the tip and the scalar operator at infinity.
We also generalize the front properties to holographic models in higher dimensional AdS soliton spacetime.
Moreover, we examine effects of the scalar mass on stability of phase transitions with numerical methods.
With $m^{2}\in (-0.38,0]$, we arrive at the classical second order insulator/superconductor phase transition.
Surprisingly, there is no stable superconducting phases in cases of $m^{2}\in (-1,-0.38]$.
In other words, superconductor
only exists in a certain range of the scalar mass in the (2+1)-dimensional AdS soliton spacetime,
which is very different from properties in other spacetime.

\end{abstract}

\maketitle
\newpage
\vspace*{0.2cm}

\section{Introduction}

The AdS/CFT correspondence provides us a novel method to analyze a strongly
interacting gauge field theory with weakly coupled AdS gravity.
It claims that a d-dimensional conformal field theory on the boundary
is related to (d+1)-dimensional AdS spacetime in the bulk \cite{Maldacena,S.S.Gubser-1,E.Witten}.
According to this correspondence, holographic metal and superconductors transition systems have been constructed
in the background of AdS black hole using
$AdS_{d+1}/CFT_{d}$ correspondences with $d=2,3,4\cdots$ \cite{SC,GM,J. Ren}.
These gravitational duals have attracted a lot of attentions for their
potential applications in condensed matter physics, for references see
\cite{S.A. Hartnoll}-\cite{Y. Liu}.

Besides the AdS black hole, another gravitational configuration AdS soliton
with the same boundary topology was
obtained by a double Wick rotation of the AdS Schwarzschild black hole \cite{GC}.
Recently, the holographic insulator and superconductor transition model
was also established in the background of five dimensional AdS soliton \cite{TS}.
It was shown that if one include a scalar field and a Maxwell field coupled in the AdS soliton,
there is second order phase transition at a critical chemical potential,
above which the non-zero scalar operator turns on.
The analysis was performed in the probe limit or
the backreaction of the matter fields on the metric was neglected. Considering the backreaction of the matter fields on
the soliton background, a first order phase transition was observed
when the backreaction is heavy enough \cite{GB}. For other progress, please see \cite{YP,cai-2,Yan Peng-1}. The
gravity duals have also been investigated in four dimensional AdS soliton \cite{YB}.
Up to now, holographic superconductors have been constructed in four dimensional or higher dimensional AdS soliton.
The existence of a lower dimensional $AdS_{3}/CFT_{2}$ dictionary depends upon the string theory.
But in fact, $AdS_{3}/CFT_{2}$ correspondence is proved to work well in studying the holographic superconductor
in the three dimensional AdS BTZ black hole \cite{J. Ren,YL,NL,PC,DKM,HBZ}.
So it is interesting to extend the discussion to three dimensional AdS soliton
to examine holographic properties.

Most of holographic properties were obtained based on numerical
solutions since equations of motion are nonlinear and coupled. Lately, analytical methods
were also applied to study properties of holographic phase transitions,
such as the Sturm-Liouville variational, the small parameter perturbation and the matching methods,
for references see \cite{GJ,CP,RS,DR,QB,SG,WW,WHH}.
These analytical approaches were proved to
be useful to search for critical phase transition points and also
qualitative properties. For example, it showed that
the critical chemical potential decreases as we choose a more negative mass in the background of AdS soliton \cite{LN,QP,RH}.
Since there is usually much more richer potential physics behind an exact formula,
we plan to give the critical chemical potential as a
function of the scalar mass in exact expression with fully analytical methods in this work.

This work is organized as follows. In section II, we construct
a holographic superconductor model in the three dimensional AdS soliton background in the probe limit. Part A of section III
is devoted to the study of holographic phase transitions by analytical methods.
In part B of section III, we further explore properties of phase transitions with numerical superconducting solutions.
We summarize our main results in the last section.

\section{Equations of motion and boundary conditions}

We begin with the simple Abelian Higgs model in $AdS_{3}$ spacetime containing a
Maxwell field and a scalar field coupled in the form
\begin{eqnarray}\label{lagrange-1}
\mathcal{L}=-\frac{1}{4}F^{\mu\nu}F_{\mu\nu}+|\nabla\psi-iqA\psi|^{2}-m^{2}|\psi|^{2},
\end{eqnarray}
where $A_{\mu}$ and $\psi$ are the Maxwell field and charged scalar field respectively.
$m^2$ is the mass of the scalar field, which plays an essential role in the condensation.
And q is the charge of the scalar field coupled to the Maxwell field.

We choose the background of the standard three dimensional AdS soliton as \cite{GC}
\begin{eqnarray}\label{AdSBH}
ds^{2}&=l^{2}\frac{dr^{2}}{f(r)}-r^2dt^{2}+f(r)d\chi^{2},
\end{eqnarray}
where $f(r)=r^{2}-r_{0}^{2}$ and $l$ is the radius of AdS spactime. In order to get rid of the conical
singularity $r_{0}$, we impose a period $\chi\sim\chi+\frac{2\pi L}{r_{0}}$
on the coordinate $\chi$ \cite{GC}.

The other matter fields of interest are as follows
\begin{eqnarray}\label{AdSBH}
A=\phi(r)dt, ~~~~~~\psi=\psi(r)e^{-i\omega t}.\nonumber
\end{eqnarray}
Using the symmetry $\psi\rightarrow \psi e^{i a t}, \phi\rightarrow \phi +\frac{a}{q}$, we set $\omega=0$.
With these assumptions, we obtain equations of motion from the action
\begin{eqnarray}\label{BHpsi}
\psi''+\left(\frac{f'}{f}+\frac{1}{r}\right)\psi'+\frac{q^2l^2}{r^2f}\phi^{2}\psi-\frac{m^{2}l^2}{f}\psi=0,
\end{eqnarray}
\begin{eqnarray}\label{BHphi}
\phi''+\left(\frac{f'}{f}-\frac{1}{r}\right)\phi'-\frac{2q^2l^2\psi^{2}}{f}\phi=0.
\end{eqnarray}
Since the equations are coupled and nonlinear, we solve these equations
by numerically integrating them from the tip $r_{0}$ out to the infinity.

In the following numerical calculation, we scale $l$ unity with the symmetries

\begin{eqnarray}\label{symmetryBH}
l \rightarrow al,~~~~~~~~(t,\chi)\rightarrow
~a(t,\chi),~~~~~~~q\rightarrow
q/a,~~~~~~~m\rightarrow m/a,
\end{eqnarray}
which leads to $ds^2\rightarrow a^2ds^2$.
Multiplying the density (1) by $q^{2}$ and preforming the rescalings $\psi\rightarrow\psi/q$ and $\phi\rightarrow\phi/q$,
we can take $q=1$ without loosing generality in the following discussion.

These equations are also
invariant under the scaling
\begin{eqnarray}\label{symmetryBH}
r \rightarrow ar,~~~~~~~~(t,\chi)\rightarrow
~(t,\chi)/a,~~~~~~~\phi\rightarrow
a\phi,~~~~~~~f\rightarrow a^2f
\end{eqnarray}
which can be used to set $r_{0}=1$ and leave the metric $ds^2$ unchanged.

We choose $m^{2}$ above the BF
bound $m_{BF}^{2}=-\frac{(d-1)^2}{4}=-1$, where $d$ is the dimension of the spacetime \cite{P. Breitenlohner}.
Near the AdS boundary $(r\rightarrow \infty)$, the asymptotic behaviors of the scalar and Maxwell fields are
\begin{eqnarray}\label{InfBH}
\psi=\frac{\psi_{-}}{r^{\lambda_{-}}}+\frac{\psi_{+}}{r^{\lambda_{+}}}+\cdot\cdot\cdot,\
\phi=\mu+\rho\ln(r)+\cdot\cdot\cdot, \ \
\end{eqnarray}
where $\lambda_{\pm}=1\pm\sqrt{1+m^{2}}$. $\mu$ and $\rho$ are interpreted as
the chemical potential and charge density in the dual theory respectively.
We will fix $\psi_{-}=0$ and the phase transition in the dual CFT is described by the
operator $\psi_{+}=<O_{+}>$ in the following discussion.

\section{The scalar condensation in AdS soliton}

\subsection{Analytical methods in holographic phase transitions}

It was revealed in Ref.\cite{TS} that when the chemical potential
exceeds a critical value, the condensation will set in.
This procedure was interpreted as insulator/superconductor transitions.
The critical value of the chemical potential is the turning point of a superconductor phase transition.
In this part, we use analytical methods to investigate the critical
value of the chemical potential of holographic
superconductors in the probe limit.
As usual, we firstly introduce a new variable $z=\frac{1}{r}$. Then equations of the scalar and Maxwell fields can be written as
\begin{eqnarray}\label{BHpsi}
\psi''+\left(\frac{f'}{f}+\frac{1}{z}\right)\psi'+\frac{\phi^{2}}{z^2f}\psi-\frac{m^{2}}{z^4f}\psi=0,
\end{eqnarray}
\begin{eqnarray}\label{BHphi}
\phi''+\left(\frac{f'}{f}+\frac{3}{z}\right)\phi'-\frac{2\psi^{2}}{z^4f}\phi=0.
\end{eqnarray}

At the phase transition points, $\psi(z)=0$. So equation (9) can be set as
\begin{eqnarray}\label{BHphi}
\phi''+\left(\frac{f'}{f}+\frac{3}{z}\right)\phi'=0.
\end{eqnarray}

We choose a simple solution $\phi=\mu$ at the phase transition point,
where $\mu$ is the chemical potential.
When the first scalar operator is fixed as $<O_{-}>=0$,
the second scalar operator
$\varepsilon=<O_{+}>$ is small close to the critical point.
Following the perturbation scheme in \cite{RH}, we introduce the scalar operator as an expansion parameter
\begin{eqnarray}\label{BHpsi}
\varepsilon=<O_{+}>.
\end{eqnarray}
Note that, in the perturbation method and close to the critical point, our interest is in the
solutions with small charge density or $\rho\thickapprox 0$. We choose the expansions as
\begin{eqnarray}\label{BHpsi}
\mu=\mu_{0}+\varepsilon^{2}\mu_{1}+\cdots.
\end{eqnarray}
These assumptions imply the relation
$<O_{+}>\thicksim(\mu-\mu_{c})^{\beta}$ with $\beta=\frac{1}{2}$.
The relations were already analytically achieved in the s-wave holographic insulator/superconductor
phase transition in higher dimensions \cite{LN,RH}.
This critical exponent $\beta=\frac{1}{2}$ for the condensation
value and $\mu-\mu_{c}$ is the same as the mean field theory and also in accordance with our numerical data in part B.
By fitting the numerical date in the next part, relations of the scalar operator
with respect to the critical chemical potential near phase transition points are derived as
$<O_{+}>\thickapprox 3.5\sqrt{\mu-\mu_{c}}$ for $m^2=0$ and $<O_{+}>\thickapprox 4.0 \sqrt{\mu-\mu_{c}}$
for $m^2=-\frac{1}{10}$.

We also have $\psi\thicksim \varepsilon$, for reference see \cite{CP}. Putting (12) into (8) and considering the 1-order of $\varepsilon$, we get
\begin{eqnarray}\label{BHpsi}
\psi''+\left(\frac{f'}{f}+\frac{1}{z}\right)\psi'+\frac{\mu_{0}^{2}}{z^2f}\psi-\frac{m^{2}}{z^4f}\psi=0.
\end{eqnarray}
Choosing $m^2=0$, we find the solution of (13) as
\begin{eqnarray}\label{BHpsi}
\psi=C_{1}M[z]+z^2 C_{2}H[z].
\end{eqnarray}
where
\begin{eqnarray}\label{BHpsi}
M[z]=MeijerG[\{\{0,0\},\{1-\mu_{0}/2,(2+\mu_{0})/2\}\},\{\{0,1\},\{0,0\}\},z^2],\\
H[z]=Hypergeometric2F1[1-\mu_{0}/2,1+\mu_{0}/2,2,z^2].~~~~~~~~~~~~~~~~~~~~
\end{eqnarray}

When $m^{2}\neq 0$, the solution of (13) is in the form
\begin{eqnarray}\label{BHpsi}
\psi=z^{1-\sqrt{1+m^2}} D_{1} H_{1}[z] + z^{1+\sqrt{1+m^2}} D_{2} H_{2}[z].
\end{eqnarray}
where
\begin{eqnarray}\label{BHpsi}
H_{1}[z]=Hypergeometric2F1[\frac{1}{2}-\frac{\mu_{0}}{2}-\frac{\sqrt{1+m^2}}{2},\frac{1}{2}+\frac{\mu_{0}}{2}-\frac{\sqrt{1+m^2}}{2},1-\sqrt{1+m^2}], z^2],\\
H_{2}[z]=Hypergeometric2F1[\frac{1}{2}-\frac{\mu_{0}}{2}+\frac{\sqrt{1+m^2}}{2},\frac{1}{2}+\frac{\mu_{0}}{2}+\frac{\sqrt{1+m^2}}{2},1+\sqrt{1+m^2}], z^2].
\end{eqnarray}

$H[z]$, $H_{1}[z]$ and $H_{2}[z]$ are the Gauss Hypergeometric function 2F1 and $M[z]$ is the  Meijer G-function.
$C_{1}$, $C_{2}$, $D_{1}$ and $D_{2}$ are integration constants.
Considering the boundary condition $\psi= <O_{+}>z^{1+\sqrt{1+m^2}}$ around $z=0$, we set $C_{1}=0,D_{1}=0$.
On this assumption, we have $\psi=D z^{1+\sqrt{1+m^2}} H_{2}[z]$ with $D$ as the constant.
In order to make $H_{2}[z]$ finite as $z\rightarrow 1$, we simply take $\frac{1}{2}-\frac{\mu_{0}}{2}+\frac{\sqrt{1+m^2}}{2}=0, -1, -2, -3,\cdots$.
In particular, we choose the minimal $\mu_{c}=\mu_{0}=1+\sqrt{1+m^2}$ as the critical chemical potential and the larger ones
correspond to higher energy states, which is exactly supported by numerical results.
From the formula $\mu_{c}=1+\sqrt{1+m^2}$,
it is clear that $\mu_{c}$ becomes smaller as we choose a more negative scalar mass $m^{2}$.
It means the more negative mass makes the condensation more easier to happen.
Since $H_{2}[z]\equiv1$ for $\mu_{c}=\mu_{0}=1+\sqrt{1+m^2}$, we find the approximate expression for the scalar
field as $\psi=<O+> z^{1+\sqrt{1+m^2}}$ around the phase transition points.
That leads to the correspondence $\psi_{0}=<O_{+}>$, where $\psi_{0}$ is the value of the scalar field $\psi$ at the tip.
Here we have simply related the value $\psi_{0}$ at the tip to the operator $<O_{+}>$ in the infinity boundary theory.
We will further examine this property in part B with numerical solutions.

In the following, we will show that our formula is astonishing precise.
We compare our analytical and numerical data in Table I. The first column is the critical chemical potential obtained
by numerical shooting method and the second column represents the critical chemical potential from our
analytical formula. The last column is with the corresponding scalar mass.
We integrate the equations in the range of $[\frac{999}{1000},\frac{1}{1000}]$
with a small value $\psi_{0}=\frac{1}{1000}$ at the tip to get our numerical data.
Considering the facts that we have chosen $r=1000$ as the infinity boundary and the
numerical method itself also has computing error, we state that the expression $\mu_{c}=1+\sqrt{1+m^2}$ is just the right form
instead of an approximation formula.

\renewcommand\arraystretch{1.7}
\begin{table} [h]
\centering
\caption{The critical chemical potential together with the scalar mass in three dimensions}
\label{address}
\begin{tabular}{|>{}c|>{}c|>{}c|}
\hline
$~~~~~~~\mu_{c}(Numerical)~~~~~~~$ & ~~~~~~~$\mu_{c}=1+\sqrt{1+m^2}$~~~~~~~ & ~~~~~~~$m^2$~~~~~~~\\
\hline
~~~~~~~2.000000~~~~~~~ & ~~~~~~~2~~~~~~~ & ~~~~~~~0~~~~~~~\\
\hline
~~~~~~~1.948684~~~~~~~ & ~~~~~~~1.948683~~~~~~~ & ~~~~~~~$-\frac{1}{10}$~~~~~~~\\
\hline
~~~~~~~1.894428~~~~~~~ & ~~~~~~~1.894427~~~~~~~ & ~~~~~~~$-\frac{2}{10}$~~~~~~~\\
\hline
~~~~~~~1.836661~~~~~~~ & ~~~~~~~ 1.836660~~~~~~~ & ~~~~~~~$-\frac{3}{10}$~~~~~~~\\
\hline
~~~~~~~1.000100~~~~~~~ & ~~~~~~~ 1.000100~~~~~~~ & ~~~~~~~$-1+\frac{1}{10^8}$~~~~~~~\\
\hline
\end{tabular}
\end{table}

Mention that $\mu_{c}=1+\sqrt{1+m^2}=1+\sqrt{m^2-(-1)}$ and $m_{BF}^{2}=-1$ for three dimensions.
This formula implies that the critical chemical potential depends on the difference of $m^2-m_{BF}^{2}$.
We generalize the formula to higher dimensions
as $\mu_{c}\thickapprox a+\sqrt{m^2-m_{BF}^{2}}$, where $a$ stands for the critical chemical potential when $m^2\rightarrow m_{BF}^{2}$.
We take the background of the five-dimensional AdS soliton for example, which has been studied a lot
through numerical methods.
The critical chemical potential for different values of the scalar mass
was also calculated with S-L analytical method \cite{LN}.
For five dimensions, there is $m_{BF}^{2}=-4$.
We simply take $\mu_{c}\thickapprox a+\sqrt{4+m^2}$
and fix $a=1.400$. The first column is our numerical date, the second column is with our approximate
formula and the last column is the corresponding scalar mass.
It can be easily seen from Table II that the approximate formula works well for various sets of parameters.

\renewcommand\arraystretch{1.7}
\begin{table} [h]
\centering
\caption{The critical chemical potential together with the scalar mass in five dimensions}
\label{address}
\begin{tabular}{|>{}c|>{}c|>{}c|}
\hline
$~~~~~~~\mu_{c}(Numerical)~~~~~~~$ & ~~~~~~~$\mu_{c}\thickapprox1.400+\sqrt{4+m^2}$~~~~~~~ & ~~~~~~~$m^2$~~~~~~~\\
\hline
~~~~~~~3.404~~~~~~~ & ~~~~~~~3.400~~~~~~~ & ~~~~~~~0~~~~~~~\\
\hline
~~~~~~~2.901~~~~~~~ & ~~~~~~~2.900~~~~~~~ & ~~~~~~~$-\frac{7}{4}$~~~~~~~\\
\hline
~~~~~~~2.396~~~~~~~ & ~~~~~~~2.400~~~~~~~ & ~~~~~~~$-\frac{12}{4}$~~~~~~~\\
\hline
~~~~~~~1.888~~~~~~~ & ~~~~~~~1.900~~~~~~~ & ~~~~~~~$-\frac{15}{4}$~~~~~~~\\
\hline
~~~~~~~1.406~~~~~~~ & ~~~~~~~1.401~~~~~~~ & ~~~~~~~$-4+\frac{1}{10^3}$~~~~~~~\\
\hline
\end{tabular}
\end{table}

\subsection{Numerical results in holographic phase transitions}

After solving the differential equations numerically, we explore properties of
phase transitions and plot the scalar operator as a function of the chemical potential in the left column of
Fig. 1 with various $m^{2}$ as $m^2=0$ (upper left), $m^2=-\frac{1}{10}$ (bottom left). For the two left panels, there is a critical chemical potential $\mu_{c}$, above which there is superconducting phase and the scalar operator increases as
we choose a larger chemical potential around the phase transition points.
In addition, it is clearly from the picture that more negative scalar mass corresponds to smaller critical
chemical potential or makes the scalar condensation more easier to happen.
These properties are similar to four dimensional and higher dimensional cases in Ref.\cite{TS,YB,QP,RH}.
We also exhibit the charge density
with respect to the chemical potential in the right column.
We see that the lines are straight or $\rho\thicksim (\mu-\mu_{c})$ around the phase transition point,
which is also in agreement with former results in \cite{TS,QP,RH}.
\begin{figure}[h]
\includegraphics[width=180pt]{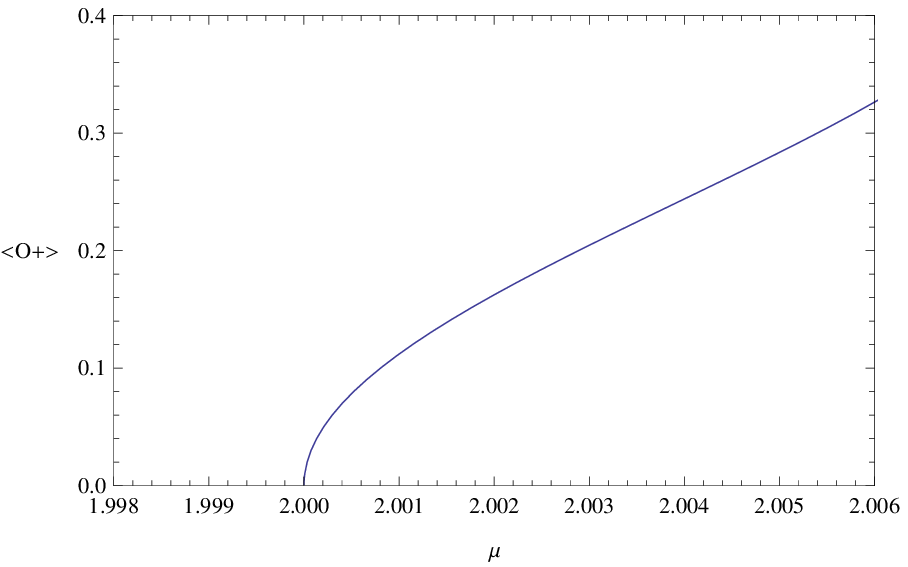}\
\includegraphics[width=170pt]{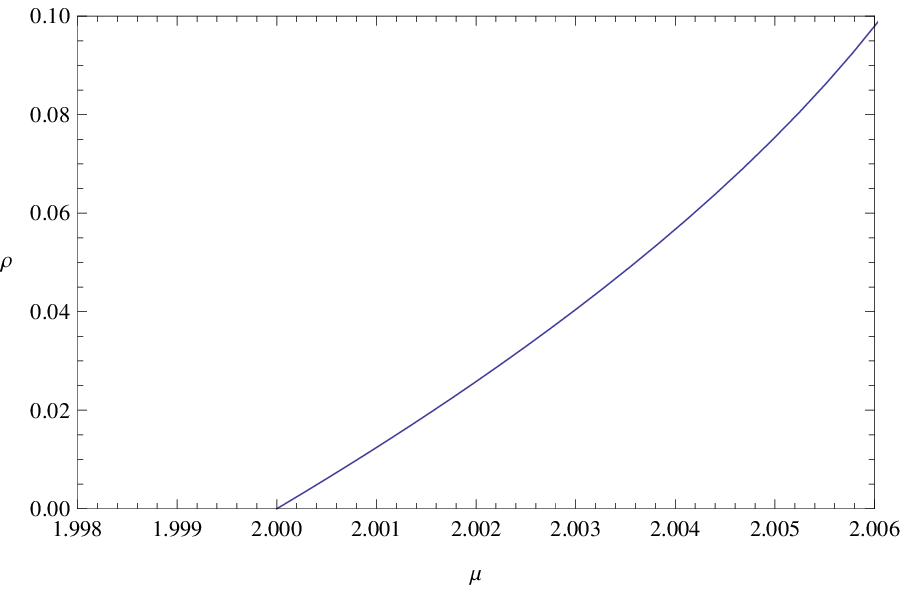}\
\includegraphics[width=180pt]{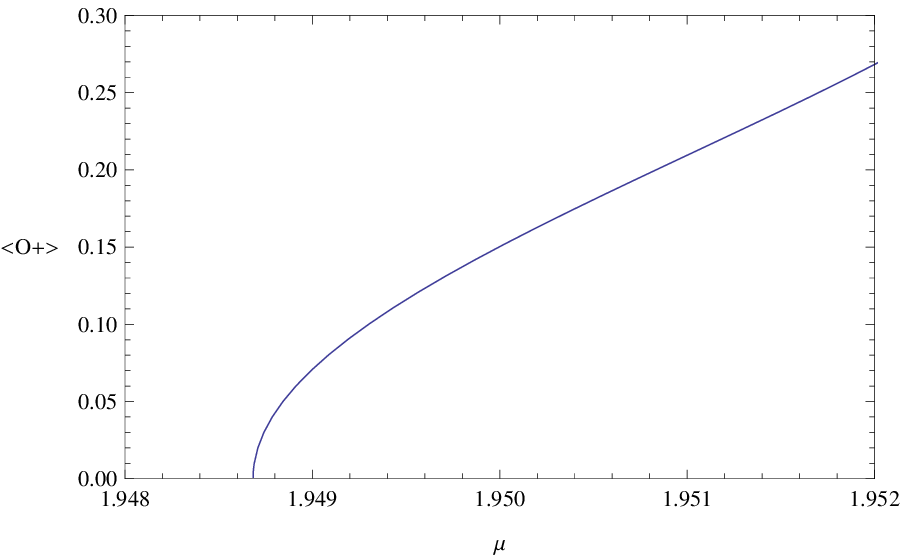}\
\includegraphics[width=170pt]{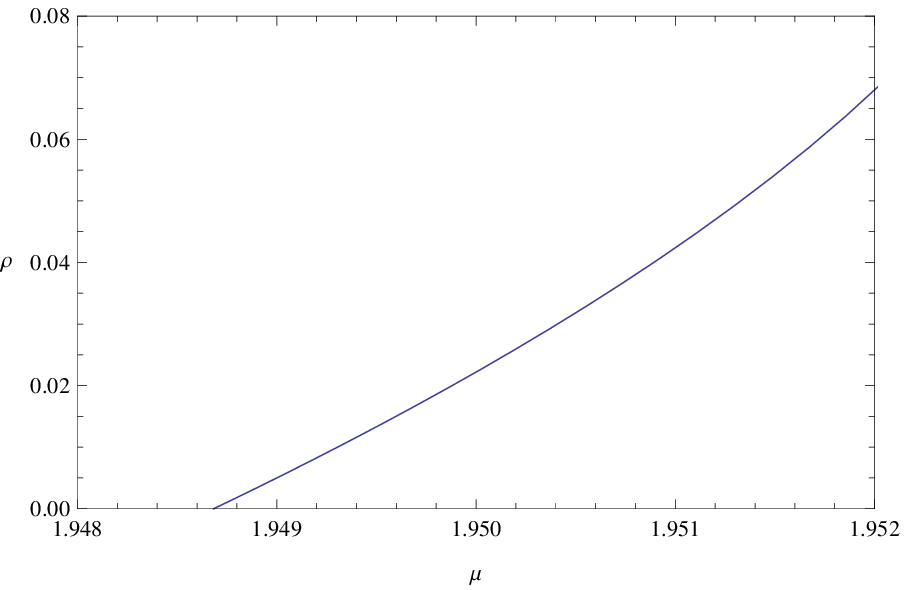}\
\caption{\label{EEntropySoliton} (Color online) The phase transitions with various $m^2$. The left column shows scalar operators as
a function of $\mu$ and the right column represents the charge density with respect to $\mu$. The two top panels are the cases
of $m^2=0$ and the two bottom panels are with $m^2=-\frac{1}{10}$.
}
\end{figure}

Surprisingly, when we choose the scalar mass $m^{2}=-\frac{5}{10}$ in Fig. 2, the scalar operator increases
as we choose a smaller chemical potential, which is not suitable to describe the insulator/superconductor system \cite{TS}.
We refer this unphysical behavior as retrograde condensation, which suggests that the superconducting phases are unstable.
This phenomenon is very different from holographic models in the background of other spacetime.
More detailed calculation shows that there is superconducting phases
for $m^{2}\in (-0.38,0]$ and when $m^{2}\in (-1,-0.38]$,
there is no superconducting phase. In summary,
the superconducting phase lives only in a certain range of the scalar mass in this model.
\begin{figure}[h]
\includegraphics[width=180pt]{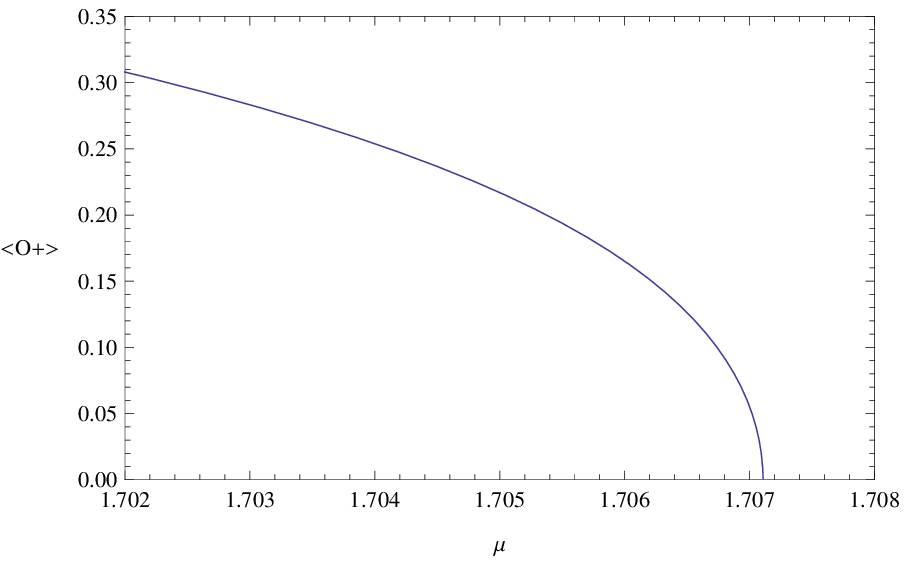}\
\includegraphics[width=170pt]{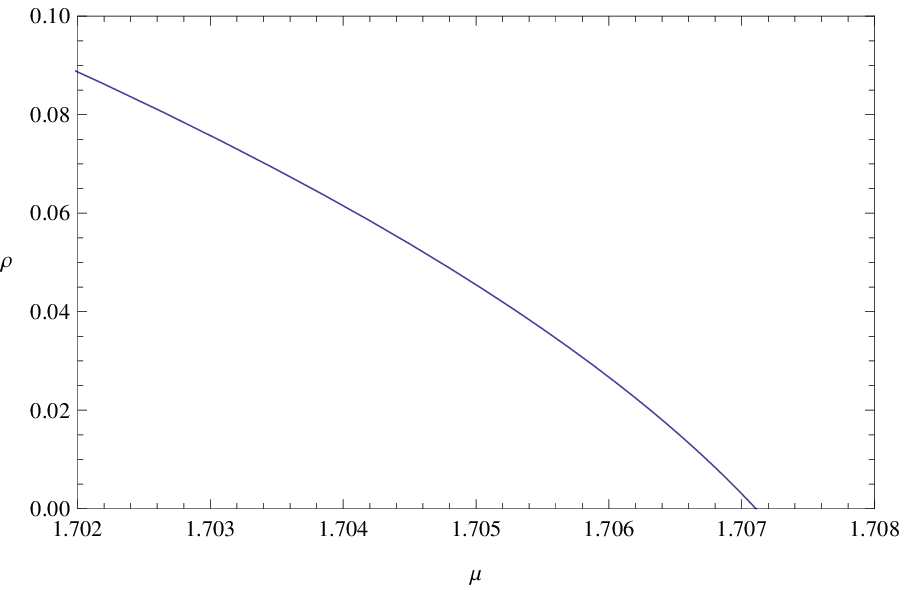}\
\caption{\label{EEntropySoliton} (Color online) The scalar condensation with $m^2=-\frac{5}{10}$. We plot scalar operators as
a function of $\mu$ in the left panel and the right panel is devoted to the charge density with respect to $\mu$.
}
\end{figure}

Now we turn to study the specific form of $\psi$ in Fig. 3. Except for the analytical approximate expression
$\psi=<O+> z^{1+\sqrt{1+m^2}}$, we can also obtain the numerical solutions
by shooting methods. We find the curves obtained from different approaches almost coincide with each other and the approximate expression
is very effective, especially for $m^2=0$. When approaching the phase transition points or
$<O+>\rightarrow 0$, the expression becomes more precise.
\begin{figure}[h]
\includegraphics[width=180pt]{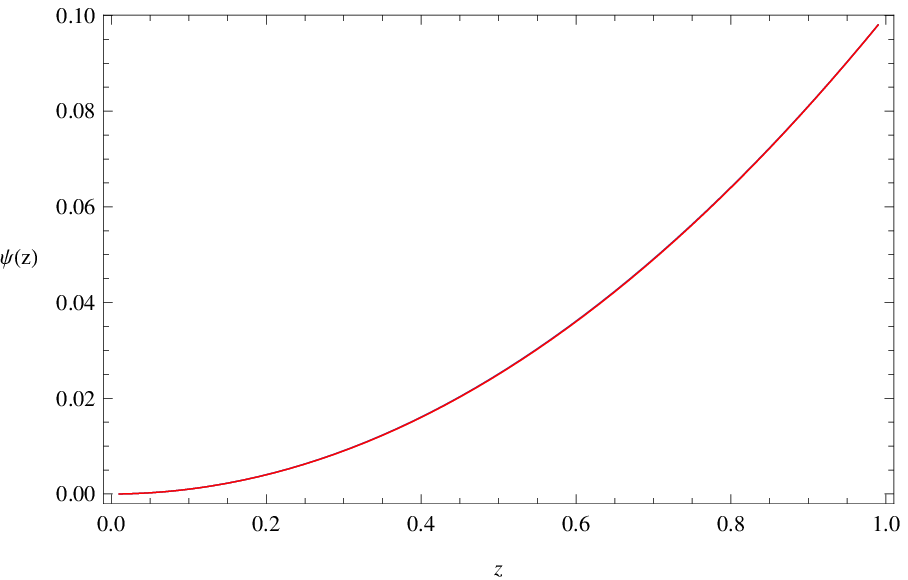}\
\includegraphics[width=180pt]{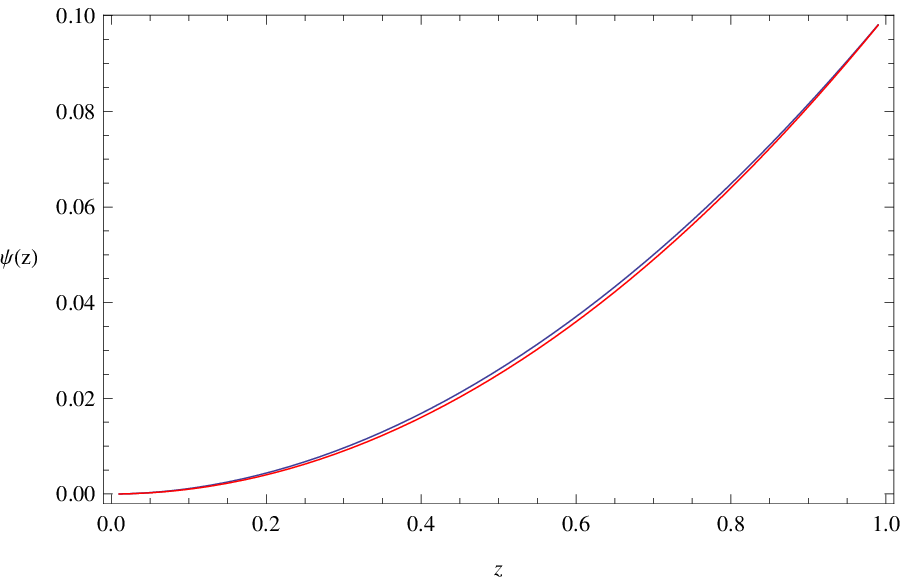}\
\caption{\label{EEntropySoliton} (Color online) We show the behaviors of the scalar field $\psi(z)$.
The blue curves represent numerical solutions with $\psi_{0}=\frac{1}{10}$
and the red curves are obtained from $\psi=<O+> z^{1+\sqrt{1+m^2}}$ with $<O+>=\frac{1}{10}$. The left panel is the case of $m^2=0$ and the right panel corresponds to $m^2=-\frac{1}{10}$ The blue curve almost coincides with the red curve in both panels.
}
\end{figure}

Inspired by the analytical results in part A, we study the scalar
operator with respect to the value of the scalar field
at the tip in Fig. 4. The analytical result $\psi_{0}=<O+>$ around the critical chemical potential
is also supported by numerical results in Fig. 4.
For example, in case of $m^2=-\frac{1}{10}$, we have $\psi_{0}=\frac{1}{100}$, $<O_{+}>=0.0100$
and $\psi_{0}=\frac{1}{10}$, $<O_{+}>=0.1006$.
For the equation (13) with coefficients depending on $z$,
this property $\psi_{0}=<O+>$ is nontrivial since $\psi_{0}$ is the value of the field and $<O_{+}>$ is the operator.
More general relation $\psi_{0}\thickapprox C <O+>$ with $C$ as a constant seems to hold in other backgrounds.
For example, our numerical data in the right panel shows that $\psi_{0}\thickapprox 0.64<O+>$
holds around the phase transition point in the five dimensional AdS soliton.
Besides AdS soliton spacetime, we also examine this correspondence in five dimensional AdS black hole,
it appears to be $\psi_{0}\thickapprox 0.22<O+>$ around the phase transition points.
\begin{figure}[h]
\includegraphics[width=180pt]{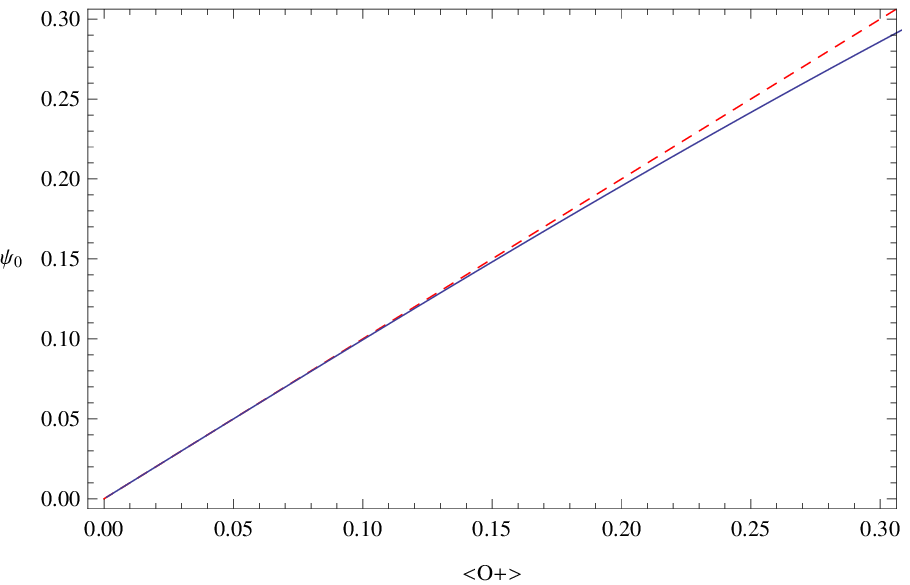}\
\includegraphics[width=180pt]{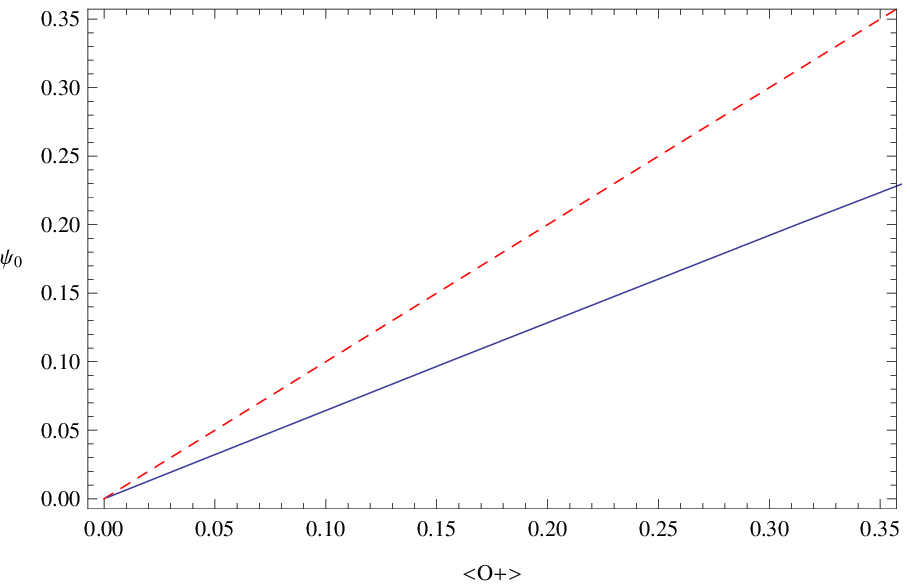}\
\caption{\label{EEntropySoliton} (Color online) We plot $\psi_{0}$ with respect to the scalar
 operator $<O+>$. The left panel is the case of $m^2=-\frac{1}{10}$ in three dimensional AdS soliton
 and the right panel is with $m^2=-\frac{15}{4}$ in five dimensional AdS soliton.
 Blue solid lines are with superconducting states. We also add the red dashed curve
 $\psi_{0}=<O+>$ as a comparison.
}
\end{figure}

\section{Conclusions}

We investigated a gravity dual in the background of three dimensional AdS soliton in the probe limit.
We studied properties of the holographic superconductors by both analytical and numerical methods.
In particular, we are the first to obtain the exact formula of the critical chemical potential as $\mu_{c}=1+\sqrt{1+m^2}$.
Since this formula is astonishing precise and seems to be just the right form,
we hope for much more richer physics explanation behind this formula in the future.
As we have showed, this formula can be generalized to higher dimensions as $\mu_{c}\thickapprox a+\sqrt{m^2-m_{BF}^{2}}$ with
the parameter $a$ depending on backgrounds.
It means more negative mass corresponds to a smaller critical chemical potential
and makes the scalar condensation more easier to happen.
In addition, the formula shows that the critical chemical potential depends
on the difference of $m^2-m_{BF}^{2}$.
Around the phase transition points, our analytical analysis showed that $\psi_{0}=<O_{+}>$,
which relates the value of the scalar field $\psi_{0}$ at the tip to the scalar operator $<O_{+}>$ at infinity.
By fitting the numerical data near the transition points, we generalized the
correspondence into other spacetime as $\psi_{0}\thickapprox C <O_{+}>$,
where $C$ is a constant depending on the metric.
With $m^{2}\in (-0.38,0]$ around the phase transition points, we arrived at the
relations $<O_{+}>\sim (\mu-\mu_{c})^{\beta}$ with $\beta=\frac{1}{2}$, which is
the same as the mean field theory implying the phase transition is of the second order .
In this case, the gravity system describes insulator/superconductor transitions
similar to higher dimensional cases.
Surprisingly, there is unusual retrograde condensation for $-1<m^2\leqslant-0.38$,
which suggests that the superconducting phases are unstable.
In other words, we have no superconductor in cases of $-1<m^2\leqslant-0.38$.
In summary, our results show that the superconducting phase lives only in a certain range of the scalar mass
in the three dimensional AdS soliton.
This property is very different from cases in other background, such as the three
dimensional BTZ black hole and other higher dimensional spacetime.

\begin{acknowledgments}

This work was supported
by the National Natural Science Foundation of China under Grant No. 11305097
and Shaanxi Province Science and Technology Department Foundation of China.
under Grant No. 2016JQ1039.

\end{acknowledgments}


\begin{thebibliography}{99}

\bibitem{Maldacena}
J.M. Maldacena,The large-N limit of superconformal field theories and supergravity, Adv. Theor. Math. Phys. {\bf 2}, 231 (1998).

\bibitem{S.S.Gubser-1}
S.S. Gubser, I.R. Klebanov, and A.M. Polyakov,Gauge theory correlators from non-critical string
theory, Phys. Lett. B {\bf
428}, 105 (1998).

\bibitem{E.Witten}
E. Witten,Anti-de Sitter space and holography, Adv. Theor. Math. Phys. {\bf 2}, 253 (1998).




\bibitem{SC}
S.A. Hartnoll, C.P. Herzog and G.T. Horowitz, Building a holographic superconductor, Phys.
Rev. Lett. 101 (2008) 031601 [arXiv:0803.3295].

\bibitem{GM}
G.T. Horowitz and M.M. Roberts, Holographic superconductors with various condensates,
Phys. Rev. D 78 (2008) 126008 [arXiv:0810.1077]

\bibitem{J. Ren}
J. Ren, One-dimensional holographic superconductor from $AdS_{3}/CFT_{2}$ correspondence,
J. High Energy Phys. 11, 055 (2010) [arXiv:1008.3904 [hep-th]].






\bibitem{S.A. Hartnoll}
S.A. Hartnoll,Lectures on holographic methods for condensed matter physics, Class. Quant. Grav. {\bf 26}, 224002 (2009).

\bibitem{C.P. Herzog}
C.P. Herzog,Lectures on Holographic Superfluidity and Superconductivity, J. Phys. A {\bf 42}, 343001 (2009).

\bibitem{G.T. Horowitz-1}
G.T. Horowitz,Introduction to Holographic Superconductors, Lect. Notes Phys. {\bf 828} 313, (2011);
arXiv:1002.1722 [hep-th].

\bibitem{E. Nakano}
E. Nakano and W.Y. Wen,Critical magnetic field in AdS/CFT superconductor, Phys. Rev. D {\bf 78}, 046004 (2008).

\bibitem{G. Koutsoumbas}
G. Koutsoumbas, E. Papantonopoulos, and G. Siopsis,Exact Gravity Dual of a Gapless
Superconductor, J. High Energy
Phys. {\bf 0907}, 026 (2009).

\bibitem{J. Sonner}
J. Sonner, A Rotating Holographic Superconductor, Phys. Rev. D {\bf 80}, 084031 (2009).

\bibitem{S.S. Gubser-2}
S.S. Gubser, C.P. Herzog, S.S. Pufu, and T. Tesileanu,Superconductors from Superstrings, Phys. Rev.
Lett. {\bf 103}, 141601 (2009).

\bibitem{Sean A. Hartnoll-3}
S.A. Hartnoll, C.P. Herzog and G.T. Horowitz,Holographic Superconductors, J. High Energy Phys.
{\bf 0812}, 015 (2008)

\bibitem{Y. Q. Liu}
Y.Q. Liu, Q.Y. Pan, and B. Wang,Holographic superconductor developed in BTZ black hole
background with backreactions, Phys. Lett. B {\bf 702}, 94 (2011).


\bibitem{J.P. Gauntlett}
J.P. Gauntlett, J. Sonner, and T. Wiseman,Holographic superconductivity in M-Theory, Phys. Rev. Lett. {\bf
103}, 151601 (2009).

\bibitem{J.L. Jing}
J.L. Jing and S.B. Chen, Holographic superconductors in the Born-Infeld electrodynamics,Phys. Lett. B {\bf 686}, 68 (2010).


\bibitem{K. Maeda}
K. Maeda, M. Natsuume, and T. Okamura,Universality class of holographic superconductors, Phys. Rev. D {\bf 79}, 126004
(2009).

\bibitem{R. Gregory}
R. Gregory, S. Kanno, and J. Soda, Holographic Superconductors with Higher Curvature Corrections, J. High Energy Phys. {\bf 0910},
010 (2009).

\bibitem{X.H. Ge}
X.H. Ge, B. Wang, S.F. Wu, and G.H. Yang, Analytical study on holographic superconductors in
external magnetic field, J. High Energy Phys. {\bf
1008}, 108 (2010).

\bibitem{Y. Brihaye}
Y. Brihaye and B. Hartmann, Holographic superconductors in 3 + 1 dimensions away from the probe
limit, Phys. Rev. D {\bf 81}, 126008 (2010).


\bibitem{CD}
C. P. Herzog, P. K. Kovtun, D. T. Son, Holographic model of superfluidity, Phys. Rev. D {\bf 79}, 066002.


\bibitem{S. Franco}
S. Franco, A.M. Garcia-Garcia, and D. Rodriguez-Gomez, A general class of holographic
superconductors, J. High Energy Phys. {\bf 1004}, 092 (2010).



\bibitem{YL}
 Yan Peng, Yunqi Liu, A general holographic metal/superconductor phase transition model, JHEP02(2015)082.


\bibitem{Y. Liu}
Y. Liu, Q. Pan and B. Wang, Holographic superconductor developed in BTZ black hole background with backreactions, Phys. Lett. B 702 (2011) 94







\bibitem{GC}
G.T. Horowitz, R.C. Myers, The AdS/CFT Correspondence and a New Positive Energy Conjecture for General Relativity,
Phys. Rev. D 59 (1998) 026005.



\bibitem{TS}
T. Nishioka, S. Ryu and T. Takayanagi, Holographic Superconductor/Insulator Transition at Zero Temperature,
JHEP 03 (2010) 131 [arXiv:0911.0962] [INSPIRE].

\bibitem{GB}
G.T. Horowitz and B. Way, Complete Phase Diagrams for a Holographic Superconductor/Insulator System, JHEP
11 (2010) 011 [arXiv:1007.3714] [INSPIRE].






\bibitem{YP}
Y. Peng, Q. Pan and B. Wang, Various types of phase transitions in the AdS soliton background, Phys. Lett. B
699 (2011) 383 [arXiv:1104.2478] [INSPIRE].

\bibitem{cai-2}
R.G. Cai, S. He, L. Li, and L.F. Li, Entanglement Entropy and Wilson Loop in $St\ddot{u}ckelberg$ Holographic
Insulator/Superconductor Model, J. High Energy Phys. {\bf
1210}, 107 (2012); arXiv:1209.1019 [hep-th].

\bibitem{Yan Peng-1}
Yan Peng, Qiyuan Pan, Holographic entanglement entropy in general holographic superconductor models,JHEP 06(2014)011.


\bibitem{YB}
Yves Brihaye, Betti Hartmann,Holographic superfluid/fluid/insulator phase transitions in 2+1 dimensions,
Phys.Rev.D83:126008,2011.






\bibitem{YL}
Yunqi Liu, Qiyuan Pan, Bin Wang, Holographic superconductor developed in BTZ black hole background with backreactions,
Phys. Lett. B.2011.06.062.

\bibitem{NL}
Nima Lashkari, Holographic Symmetry-Breaking Phases in $AdS_{3}/CFT_{2}$, JHEP11(2011)104.

\bibitem{PC}
Pankaj Chaturvedi, Gautam Sengupta, Rotating BTZ Black Holes and One Dimensional Holographic Superconductors,
Phys. Rev. D 90, 046002 (2014).

\bibitem{DKM}
Davood Momeni, Kairat Myrzakulov, Ratbay Myrzakulov, Phase transition via entanglement entropy in $AdS_{3}/CFT_{2}$ superconductors,
arXiv:1602.08718.

\bibitem{HBZ}
Hua Bi Zeng, Yu Tian, Zhe Yong Fan, Chiang-Mei Chen, Nonlinear Transport in a Two Dimensional Holographic Superconductor,
arXiv:1604.08422









\bibitem{GJ}
G. Siopsis and J. Therrien, J. High Energy Phys. 05, 013 (2010).

\bibitem{CP}
C.P. Herzog, An Analytic Holographic Superconductor, Phys. Rev. D {\bf 81}, 126009 (2010).

\bibitem{RS}
R. Gregory, S. Kanno, and J. Soda, J. High Energy Phys. 10, 010 (2009).


\bibitem{DR}
D. Momeni, R. Myrzakulov, L. Sebastiani, M. R. Setare, Int.J.Geom.Meth.Mod.Phys. 12 (2015), arXiv:1210.7965.

\bibitem{QB}
Qiyuan Pan, Jiliang Jing, Bin Wang,Analytical investigation of the phase transition
between holographic insulator and superconductor in Gauss-Bonnet gravity,JHEP 11 (2011) 088.


\bibitem{SG}
Sunandan Gangopadhyay,Analytic study of properties of holographic superconductors away from the probe limit,
Physics Letters B 724 (2013) 176-181.

\bibitem{WW}
Wen-Yu Wen, Mu-Sheng Wu, Shang-Yu Wu,A Holographic Model of Two-Band Superconductor,Phys. Rev. D 89, 066005 (2014).

\bibitem{WHH}
Wung-Hong Huang,Analytic Study of First-Order PhaseTransition in Holographic Superconductor and Superfluid,
Int. J. Mod. Phys. A 28 (2013).


\bibitem{LN}
Lukasz Nakonieczny, Marek Rogatko, Karol.I.Wysokinski,
Analytic investigation of holographic phase transitions influenced by dark matter sector,Phys.Rev.D92, 066008 (2015).

\bibitem{QP}
Qiyuan Pan, Bin Wang, Eleftherios Papantonopoulos, Jeferson de Oliveira, A.B. Pavan,
Holographic Superconductors with various condensates in Einstein-Gauss-Bonnet gravity,
Phys.Rev.D81:106007,2010.

\bibitem{RH}
Rong-Gen Cai, Huai-Fan Li, Hai-Qing Zhang,Analytical Studies on Holographic Insulator/Superconductor Phase Transitions,
Phys.Rev.D83:126007,2011.








\bibitem{P. Breitenlohner}
 P. Breitenlohner and D.Z. Freedman, Positive energy in Anti-de Sitter backgrounds and gauged
extended supergravity, Phys. Lett. B {\bf 115}, 197 (1982).




















































































\end{thebibliography}
\end{document}